\documentclass[fleqn,usenatbib]{mnras}

\usepackage[T1]{fontenc}

\DeclareRobustCommand{\VAN}[3]{#2}
\let\VANthebibliography\thebibliography
\def\thebibliography{\DeclareRobustCommand{\VAN}[3]{##3}\VANthebibliography}

\usepackage{graphicx}	
\usepackage{amsmath}	
\usepackage{amssymb}	

\usepackage{newtxtext,newtxmath}

\newcommand{\Msun}{\,\rmn{M}_{\sun}}

\newcommand{\Gyr}{\,\rmn{Gyr}}
\newcommand{\Myr}{\,\rmn{Myr}}

\newcommand{\kpc}{\,\rmn{kpc}}

\newcommand{\Mag}{\,\rmn{mag}}
\newcommand{\Arcsec}{\,\rmn{arcsec}}

\newcommand{\K}{\,\rmn{K}}

\newcommand{\kappaco}{\kappa_\mathrm{co,rot}}
\newcommand{\Reff}{R_\mathrm{eff}}

\title[EAGLE galaxy age gradients]{The age gradients of galaxies in EAGLE: outside-in quenching as the origin of young bulges in cluster galaxies}

\author[J. Pfeffer et al.]{
Joel Pfeffer,$^{1}$\thanks{E-mail: joel.pfeffer@icrar.org (JP)}
Kenji Bekki,$^{1}$
Warrick J. Couch,$^{2}$
B\"{a}rbel S. Koribalski,$^{3,4}$
Duncan A. Forbes$^{2}$
\\
$^{1}$International Centre for Radio Astronomy Research (ICRAR), M468, University of Western Australia, 35 Stirling Hwy, Crawley, WA 6009, Australia \\
$^{2}$Centre for Astrophysics \& Supercomputing, Swinburne University of Technology, Hawthorn VIC 3122, Australia \\
$^{3}$Australia Telescope National Facility, CSIRO Astronomy and Space Science, P.O. Box 76, Epping, NSW 1710, Australia \\
$^{4}$School of Science, Western Sydney University, Locked Bag 1797, Penrith, NSW 2751, Australia 
}

\date{Accepted 2021 December 31. Received 2021 December 17; in original form 2021 November 5}

\pubyear{2021}

\begin{document}
\label{firstpage}
\pagerange{\pageref{firstpage}--\pageref{lastpage}}
\maketitle

\begin{abstract}
Many disc galaxies in clusters have been found with bulges of similar age or younger than their surrounding discs, at odds with field galaxies of similar morphology and their expected inside-out formation.
We use the EAGLE simulations to test potential origins for this difference in field and cluster galaxies.
We find, in agreement with observations, that on average disc-dominated field galaxies in the simulations have older inner regions, while similar galaxies in groups and clusters have similarly aged or younger inner regions.
This environmental difference is a result of outside-in quenching of the cluster galaxies.
Prior to group/cluster infall, galaxies of a given present-day mass and morphology exhibit a similar evolution in their specific star formation rate (sSFR) profiles. 
Post-infall, the outer sSFRs of group and cluster galaxies significantly decrease due to interstellar medium stripping, while the central sSFR remains similar to field galaxies. Field disc galaxies instead generally retain radially increasing sSFR profiles.
Thus, field galaxies continue to develop negative age gradients (younger discs), while cluster galaxies instead develop positive age gradients (younger bulges). 
\end{abstract}




\section{Introduction}

Lenticular (S0) galaxies were originally defined as an `intermediate' class between ellitpical and spiral galaxies \citep{Hubble_26}, featuring a discy morphology (like spirals) but without strong spiral structure.
While spiral galaxies dominate low-density regions, S0 galaxies are more numerous in galaxy clusters \citep{Hubble_and_Humason_31, Oemler_74, Dressler_80, Postman_and_Geller_84}. 
Similarly, the fraction of S0s galaxies in these dense environments increases towards lower redshift, while the fraction of spirals decrease \citep{Butcher_and_Oemler_78, Dressler_et_al_97, Couch_et_al_98, Fasano_et_al_00, Desai_et_al_07, Just_et_al_10}. 
Both facts suggest an interpretation where spiral galaxies are transformed into S0 galaxies within galaxy clusters through processes such as ram pressure stripping, starvation, harassment and mergers \citep[e.g.][]{Spitzer_and_Baade_51, Gunn_and_Gott_72, Larson_et_al_80, Moore_et_al_96, Moore_et_al_99, Bekki_98, Bekki_09, Bekki_and_Couch_11, Querejeta_et_al_15, Merluzzi_et_al_16}.

The stellar populations of the galaxies may offer clues into their transformation processes. 
The inside-out formation of massive ($M_\ast \gtrsim 10^{10} \Msun$) field disc galaxies is well studied in the literature.
At both low ($z \sim 0$) and high ($z \sim 1$-$2$) redshifts, they typically exhibit redder, older centres with bluer, younger discs \citep[e.g.][]{Terndrup_et_al_94, de_Jong_96, Bell_and_de_Jong_00, MacArthur_et_al_04, Wang_et_al_11, Perez_et_al_13, Gonzalez_Delgado_et_al_14, Liu_et_al_16, Garcia-Benito_et_al_17, Goddard_et_al_17, Breda_and_Papaderos_18}, specific star formation rates (sSFR) that increase at larger radii \citep[e.g.][]{Munoz-Mateos_et_al_07, Tacchella_et_al_15, Nelson_et_al_16} and sizes that were smaller at higher redshift \citep[e.g.][]{Ferguson_et_al_04, Buitrago_et_al_08, van_Dokkum_et_al_13}. 
Massive field or group S0 galaxies also tend to show similar negative age gradients \citep{Fraser-McKelvie_et_al_18, Deeley_et_al_20, Dominguez_Sanchez_et_al_20, Johnston_et_al_21}.

Unlike their higher-mass counterparts, low-mass galaxies ($M_\ast \lesssim 10^{10} \Msun$; in all environments) generally show flat (self-similar formation) or positive (``outside-in'' formation) colour, age and sSFR profiles \citep[e.g.][]{van_Zee_01, MacArthur_et_al_04, Perez_et_al_13, Gonzalez_Delgado_et_al_14, Liu_et_al_16, Liu_et_al_17, Nelson_et_al_16, Wang_et_al_17, Breda_and_Papaderos_18, Fraser-McKelvie_et_al_18, Breda_et_al_20, Dominguez_Sanchez_et_al_20}.
Signatures of outside-in formation in dwarf galaxies may plausibly be due to stellar diffusion \citep{Papaderos_et_al_02}, or mergers \citep{Benitez-Llambay_et_al_16} or feedback from star formation \citep{Graus_et_al_19} heating the orbits of old stars to the large radii.

The inside-out formation of massive galaxies (and its absence in lower mass galaxies) may be a result of black hole feedback. Comparing simulations run both with and without active galactic nuclei (AGN) feedback, both \citet{Appleby_et_al_20} and \citet{Nelson_et_al_21} (comparing the SIMBA and IllustrisTNG simulations, respectively) found that AGN feedback was necessary to suppress the central star formation in galaxies (star formation feedback alone was not sufficient).

In the denser environment of galaxy clusters the picture is somewhat different. Both spiral and S0 galaxies in the clusters have been found to have bulges that are generally of similar age or younger than the surroundings discs \citep{Bedregal_et_al_11, Roediger_et_al_11, Roediger_et_al_12, Johnston_et_al_14, Barsanti_et_al_21}.
If the progenitors of these cluster galaxies were similar to \textit{present-day} field galaxies, then a simple fading or galaxy-wide quenching scenario for the origin of the cluster galaxies would appear to be ruled out.
The transformation process may quench the star formation in such galaxies from outside-in \citep{Bedregal_et_al_11} or result in a secondary episode of star formation in the bulge \citep{Johnston_et_al_14} to explain the observations.
Alternatively, the difference between field and cluster galaxies could be a result of a form of progenitor bias \citep{Woo_et_al_17}, where the field progenitors of the cluster galaxies were markedly different from present-day field galaxies.

In this work, we use the EAGLE simulations \citep{Schaye_et_al_15, Crain_et_al_15} to test the environmental dependence of galaxy age gradients and whether the observed positive age gradients in cluster galaxies may be a result of outside-in quenching, secondary episodes of star formation, or progenitor bias.
This paper is organised as follows.
In Section~\ref{sec:methods} we describe the EAGLE simulations and analysis of the simulated galaxies.
In Section~\ref{sec:results} we present the results on age differences in EAGLE galaxies, the dependence on mass, environment and morphology and its origin in the simulations.
Finally, we discuss and summarise the conclusions of this work in Sections~\ref{sec:discussion} and \ref{sec:summary}.

\section{Methods}
\label{sec:methods}

\subsection{EAGLE simulations}

This work analyses the galaxies from the Evolution and Assembly of GaLaxies and their Environments (EAGLE) project \citep{Schaye_et_al_15, Crain_et_al_15}. 
The EAGLE project is a suite of cosmological hydrodynamical simulations of galaxy and evolution formation in the $\Lambda$ cold dark matter ($\Lambda$CDM) cosmogony.
The simulations adopt cosmological parameters consistent with a \citet{Planck_2014_paperXVI_short} cosmology, in particular using $\Omega_\mathrm{m} = 0.307$, $\Omega_{\Lambda} = 0.693$, $\Omega_\mathrm{b} = 0.04825$, $h = 0.6777$ and $\sigma_8 = 0.8288$.
The simulations were run with a highly modified version of \textsc{Gadget 3} \citep{Springel_05}, which includes routines for radiative cooling, star formation, stellar evolution, stellar feedback, black holes (BHs) and AGN feedback \citep[see][for details]{Schaye_et_al_15}.

The feedback prescriptions (both stellar and BH) in EAGLE were calibrated to produce galaxies with realistic stellar masses and sizes and BH masses at $z \approx 0$ \citep{Crain_et_al_15}.
The EAGLE simulations are well studied and have been shown to broadly reproduce many features of the evolving galaxy populations, such as the galaxy stellar mass function \citep{Furlong_et_al_15} and sizes \citep{Furlong_et_al_17}, galaxy colours and star formation rates \citep{Furlong_et_al_15, Trayford_et_al_15, Trayford_et_al_17}, galaxy morphologies \citep{Bignone_et_al_20}, cold gas properties \citep{Lagos_et_al_15, Bahe_et_al_16, Crain_et_al_17} and circumgalactic and intergalactic absorption system properties \citep{Rahmati_et_al_15, Oppenheimer_et_al_16, Turner_et_al_16}.

Structures were identified in the simulation snapshots using the friends-of-friends \citep[FOF,][]{Davis_et_al_85} and \textsc{subfind} \citep{Springel_et_al_01, Dolag_et_al_09} algorithms.
Dark matter haloes were first identified by running the FoF algorithm on the dark matter particles with a linking length of $0.2$ times the mean interparticle separation.
Bound subhaloes (galaxies) were then identified within the FoF groups with \text{subfind}.
In each FoF halo, the subhalo containing the particle with the minimum gravitational potential is defined as the central galaxy and all others are defined as satellite galaxies.
To link galaxies between snapshots, galaxy merger trees were created using the D-Trees algorithm \citep{Jiang_et_al_14, Qu_et_al_17}.

In this work we analyse the EAGLE reference simulation (Ref-L100N1504) of a periodic volume 100 comoving Mpc on a side with $1504^3$ gas and dark matter particles. 
Initial gas and dark matter particle mass are $m_g = 1.8 \times 10^6 \Msun$ and $m_\mathrm{dm} = 9.7 \times 10^6 \Msun$, respectively, while the maximum gravitational softening length is $0.7$ proper kpc.

\subsection{Galaxy sample and analysis}

In this work we perform a qualitative comparison with observational studies, since we cannot apply the same techniques (e.g. SED fitting or Lick index analysis) and systematic differences exist between different studies \citep[e.g. due to adopted stellar population models or the adoption of light- or mass-weighted ages, as discussed by][]{Barsanti_et_al_21}.
Moreover, each study often applies unique definitions for bulges/discs or inner/outer galaxy (such as spectroscopic bulge-disc decomposition in \citealt{Johnston_et_al_14} or radial limits in \citealt{Fraser-McKelvie_et_al_18}).
With this in mind, we analyse the EAGLE galaxies in an observationally-motivated way as follows.

We first restrict the sample of simulated galaxies to those with stellar masses $M_\ast(r<30\kpc) > 10^9 \Msun$ at $z=0$, where the particles are bound to the galaxy according to the \textsc{subfind} algorithm.
This limit is similar to observed galaxy samples \citep[e.g.][]{Fraser-McKelvie_et_al_18}.
Following the galaxy merger trees \citep{Qu_et_al_17}, we also analyse the main progenitors of the $z=0$ galaxies at redshifts $0.5$, $1$ and $2$.

Next, we define an outer radius limit within which to analyse the galaxies.
Projecting the galaxies in the $x$-$y$ plane of the simulation volume (i.e. random orientations) we calculate galaxy surface brightnesses within $1 \kpc$ annuli from the centre of potential.
The luminosities of star particles were calculated using \textsc{fsps} \citep{Conroy_Gunn_and_White_09, Conroy_and_Gunn_10}, assuming simple stellar populations for each particle and no dust attenuation.
For each galaxy we calculate $R_{25}$, the radius at which the $r$-band surface brightness drops below $25 \Mag \Arcsec^{-2}$.
Galaxies with a maximum surface brightness fainter than $25 \Mag \Arcsec^{-2}$ were excluded from the sample.

\begin{figure*}
  \includegraphics[width=0.49\textwidth]{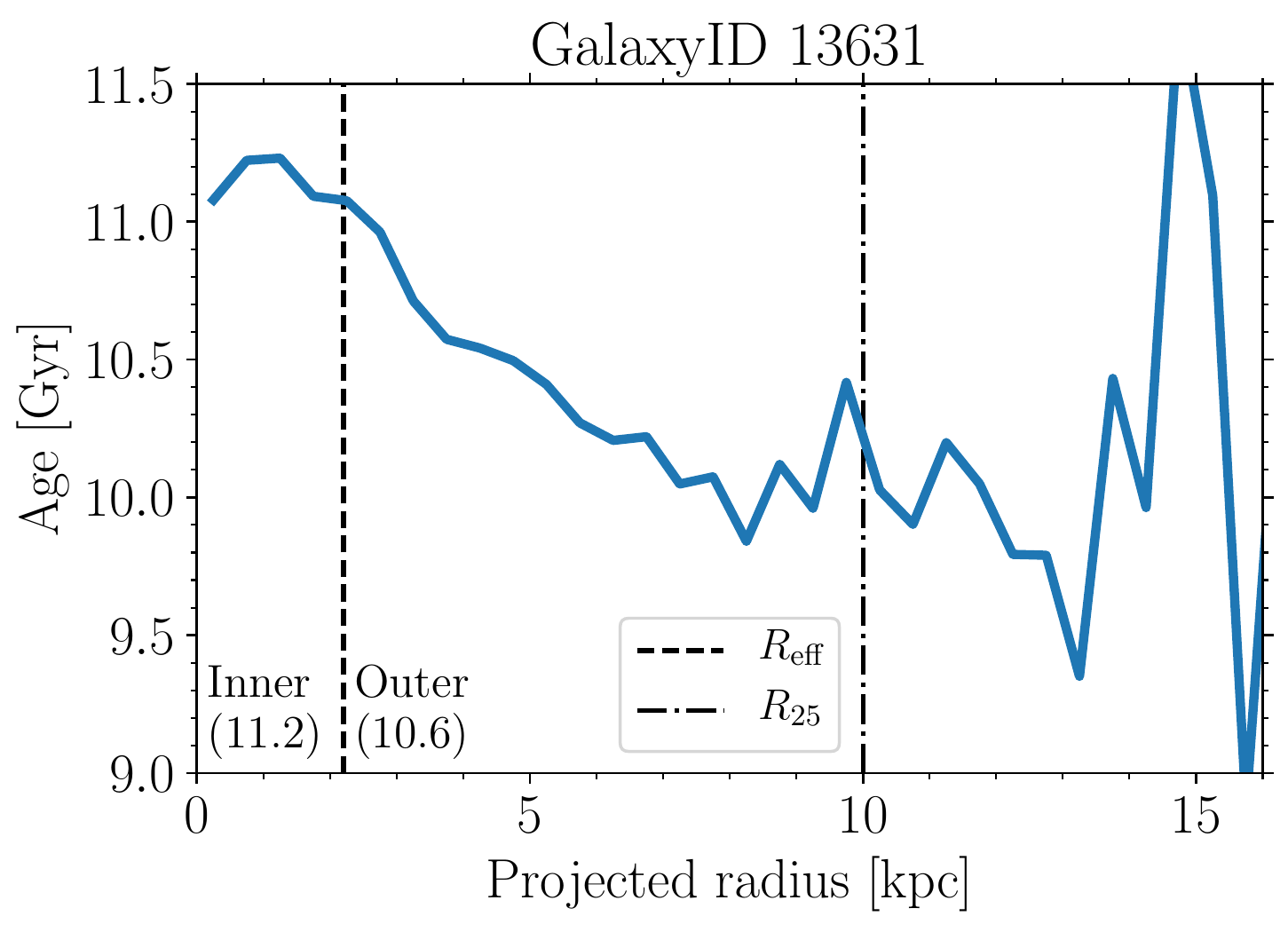}
  \includegraphics[width=0.49\textwidth]{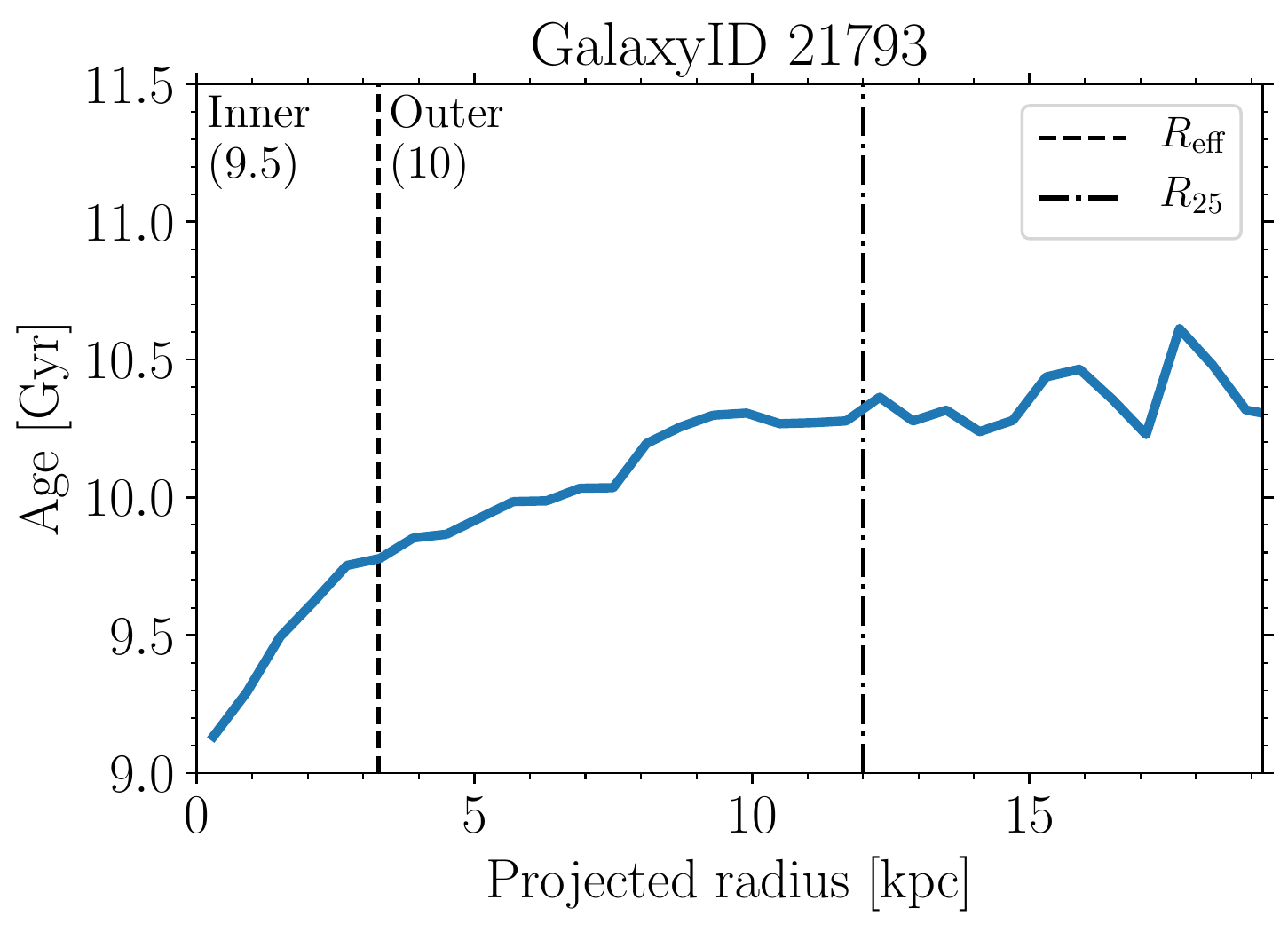}
  \caption{Two example galaxies showing negative (left) and positive (right) age gradients. The vertical lines show the $r$-band half-light radius (dashed line) and $R_{25}$ (dash-dotted line). The inner and outer mass-weighted ages (in Gyr) are shown within parentheses in each panel. Both galaxies reside the same $1.9 \times 10^{14} \Msun$ halo, have similar stellar masses ($\sim 4 \times 10^{10} \Msun$) and $\kappaco$ ($0.3$-$0.4$), and are non-star forming at $z=0$.}
  \label{fig:AgeProfs}
\end{figure*}

We define the effective radius for each galaxy ($\Reff$) as the $r$-band half-light radius with $R_{25}$ (typically $R_{25} \approx 3 R_\mathrm{eff}$).
The mass-weighted average ages and metallicities of stars were calculated in the inner ($R<\Reff$) and outer ($\Reff < R < R_{25}$) regions of each galaxy.\footnote{We also investigated fitting \citet{Sersic_63} $+$ exponential profiles and splitting the galaxies into bulge $+$ disc \citep[somewhat similar to][]{Fraser-McKelvie_et_al_18}, but found the results for age differences were consistent with dividing the galaxies at $\Reff$. Therefore we adopt the simpler method of dividing the galaxies at $\Reff$, which is also applicable for spheroidal galaxies and bulge-less galaxies (exponential discs).}
We show two example age profiles in Fig.~\ref{fig:AgeProfs}, indicating the inner and outer mass-weighted ages in each case.

To compare the environmental dependence of galaxy age gradients, we use the halo virial mass $M_{200}$ as a proxy for environment.
$M_{200}$ is defined as the total mass within $R_{200}$, the radius within which the mean density is 200 times the critical density ($3 H^2 / 8 \pi G$).
We break the galaxy sample into `field' galaxies ($M_{200} < 10^{13} \Msun$), `galaxy groups' ($10^{13} \leq M_{200} /\mathrm{M}_{\sun} < 10^{14}$) and `galaxy clusters' ($M_{200} \geq 10^{14} \Msun$).

We also investigate the effect of galaxy `morphology'. 
We use the kinematic morphology indicator $\kappaco$, the fraction of stellar kinetic energy invested in ordered co-rotation \citep[$\kappaco$ is tightly correlated with properties such as $v_\mathrm{rot} / \sigma$ and spin parameter $\lambda_\ast$;][]{Correa_et_al_17, Thob_et_al_19}.
We break the galaxy sample into `spheroids' ($\kappaco < 0.35$), `S0-like' galaxies ($0.35 < \kappaco < 0.45$) and `late-type' (disc dominated) galaxies ($\kappaco > 0.45$).
We also further break the later group into `moderately discy' ($0.45 < \kappaco < 0.6$) and `very disc-dominated' ($\kappaco > 0.6$) galaxies.
We note however that there will not be a perfect correlation between $\kappaco$ and visual morphology.
For example, a disc-dominated ($\kappaco > 0.45$) galaxy with active star formation may appear as a spiral galaxies, while a quenched galaxy with similar $\kappaco$ would appear to be an S0 galaxy.

\subsection{Star formation peak finding}
\label{sec:peakFinding}

In Section~\ref{sec:SFpeaks} we investigate the occurrence of secondary star formation peaks in galaxies. We search for star formation peaks using the method as follows:
\begin{itemize}
  \item We determine the star formation histories of the galaxies in intervals of $1 \Gyr$ width. We chose a relatively wide interval given we are not concerned with brief fluctuations (e.g. periodic episodes from star formation followed by stellar feedback) but protracted periods of quiescence followed by reinvigorated star formation.
  \item We then search for maxima and minima in the formation histories. Valid peaks are defined to occur once the star formation history has increased by more than 50 per cent from the previous minimum.
  \item Once a maximum has been achieved, the peak is defined to end once the star formation history has dropped by a factor $1/1.5$ (as before, the peak must be 50 per cent above minimum). Following a valid minimum, we may then search for the next maximum/star formation peak (if one occurs).
  \item Peaks below a mass fraction of 1 per cent are ignored due to possible noise from the coarse particle mass resolution ($\sim10^6 \Msun$) and given they contribute little to the mass budget of the galaxy.
\end{itemize}

\section{Results}
\label{sec:results}

\subsection{Inner/outer age differences}
\label{sec:dAge}

\begin{figure*}
  \includegraphics[width=\textwidth]{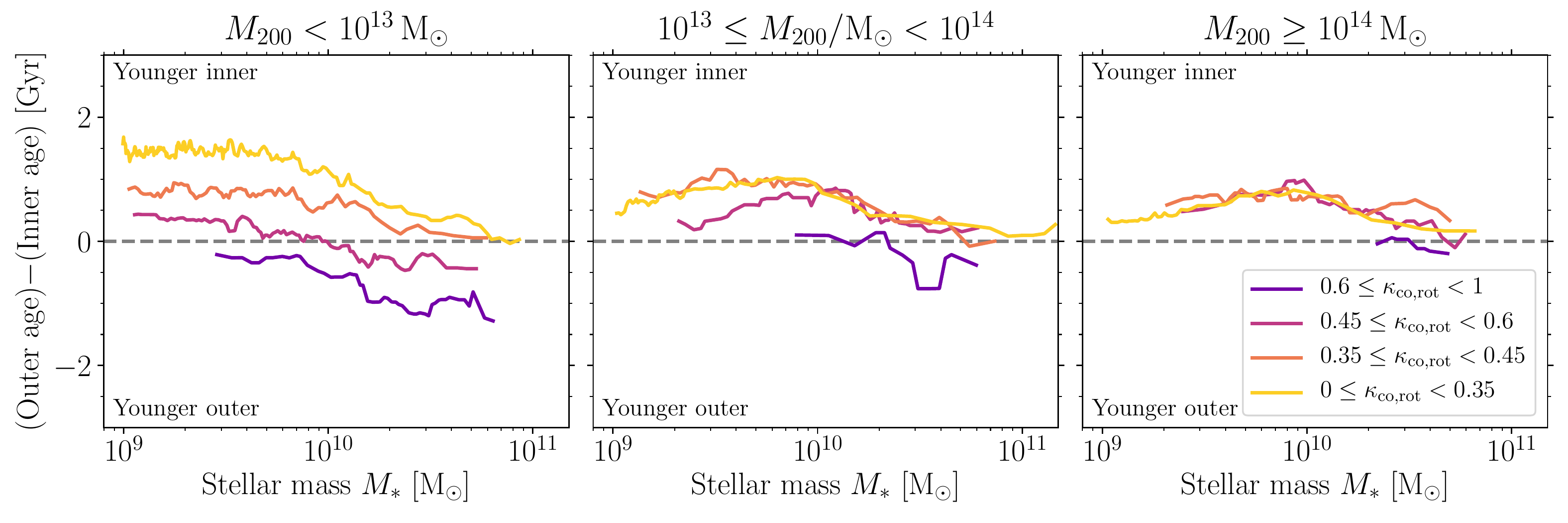}
  \caption{Outer$-$inner galaxy age differences as a function of galaxy mass, with the galaxies divided into subsamples by environment (panels) and morphology ($\kappaco$, lines coloured according to legend). The lines show running medians for each subsample as a function of galaxy mass, calculated with a fixed number of galaxies using a bin width of one tenth the sample size (with a minimum number of 20 and maximum number of 100 galaxies per bin) and incrementing the bin by one fifth the bin size for each measurement. While disc-dominated ($\kappaco > 0.45$) field galaxies show negative age differences (younger outer regions) in the median, such galaxies in groups and clusters tend to show near zero or positive age differences in groups and clusters.}
  \label{fig:dAge}
\end{figure*}

In the upper panel of Fig.~\ref{fig:dAge} we first compare the age difference between the inner and outer regions (divided at $\Reff$) of galaxies as a function of galaxy mass, galactic environment ($M_{200}$) and kinematic morphology ($\kappaco$).
For reference, in Appendix~\ref{app:colours} we also compare metallicity and colour differences for the same galaxies.
The ages and metallicities for each region (inner/outer galaxy) are both calculated as mass-weighted means.
We note that there are few low-mass, very disc-dominated galaxies ($<10^{10} \Msun$, $\kappaco > 0.6$, the median masses of such galaxies in all environments are in the range $10^{10.2}$-$10^{10.5} \Msun$).
This appears particularly evident for cluster galaxies ($M_{200} \geq 10^{14} \Msun$), relative to field galaxies ($M_{200} < 10^{13} \Msun$), which is a result of poor sampling and the binning procedure in Fig.~\ref{fig:dAge} (598, 81 and 47 field, group and cluster galaxies, respectively).
The lack of such galaxies is possibly due to the minimum galaxy size and disc thickness imposed by the polytropic equation of state used in EAGLE \citep{Schaye_et_al_15, Furlong_et_al_17} and/or mass segregation from unequal baryonic and dark matter particle masses resulting in spurious collisional heating \citep{Ludlow_et_al_19, Ludlow_et_al_21}.
However, observed galaxies on the star formation main sequence also show a decrease in spin parameter for stellar masses $<10^{10} \Msun$ \citep{Wang_et_al_20}.
In Appendix~\ref{app:variations}, we compare the age differences in two different EAGLE model variations (higher resolution model and modified AGN feedback), finding good agreement with the reference model. Therefore for the remainder of this work we focus on the reference model given the larger volume and number of galaxies.

We find that the median age differences in EAGLE galaxies are generally less than $\sim 1 \Gyr$ (with the exception of low-mass, field spheroidal galaxies), i.e. the inner and outer ages of galaxies are correlated, though the average age of galaxies scales with galaxy mass due to downsizing \citep{Bower_et_al_92, Cowie_et_al_96, Gallazzi_et_al_05}.
A correlation between inner and outer ages has been similarly found for observed galaxies \citep[e.g.][]{MacArthur_et_al_04, Gonzalez_Delgado_et_al_14, Fraser-McKelvie_et_al_18, Pak_et_al_21}, though the magnitude of the differences will of course depend on measurement details (e.g. light- versus mass-weighted ages).

For field galaxies ($M_{200} < 10^{13} \Msun$) there is a clear trend between age difference and $\kappaco$, such that very discy galaxies ($\kappaco > 0.6$) have negative age gradients (younger outer regions) while spheroidal galaxies ($\kappaco < 0.35$) have positive gradients (older outer regions). 
There is also a mild trend with galaxy mass (most pronounced for spheroidal galaxies), such that low mass galaxies predominantly have younger inner regions, while massive galaxies tend to have younger outer regions (or no age difference).
These results are in line with the trends found for observed galaxies, with low-mass galaxies ($<10^{10} \Msun$) predominantly showing ``outside-in'' formation \citep[e.g.][]{Perez_et_al_13, Fraser-McKelvie_et_al_18, Breda_et_al_20} and high-mass galaxies ($>10^{10} \Msun$) showing ``inside-out'' formation \citep[e.g.][]{MacArthur_et_al_04, Garcia-Benito_et_al_17, Fraser-McKelvie_et_al_18}.
Interestingly, the transition from positive to negative age gradients at $\sim 10^{10} \Msun$ similarly occurs for observed late-type galaxies \citep{Breda_and_Papaderos_18, Breda_et_al_20}.
The apparent outside-in formation of dwarf galaxies may be due to stellar diffusion, mergers or stellar feedback heating the orbits of old stars \citep[e.g.][]{Papaderos_et_al_02, Benitez-Llambay_et_al_16, Graus_et_al_19}.

In contrast with field galaxies, the age differences of group ($10^{13} < M_{200}/\mathrm{M}_{\sun} < 10^{14}$) and cluster ($M_{200} > 10^{14} \Msun$) galaxies predominantly show younger inner regions or little age difference. 
Again, this result is in good agreement with observational findings that late-type and S0 galaxies in clusters have bulges of similar age to or younger than their surrounding discs \citep{Bedregal_et_al_11, Roediger_et_al_11, Roediger_et_al_12, Johnston_et_al_14, Barsanti_et_al_21}.
There is less dependence on morphology in the age differences (particularly in clusters) with mainly $\kappaco > 0.6$ galaxies showing more negative age differences.
As we later discuss in Section~\ref{sec:zEvo}, the near-zero age difference for $\kappaco > 0.6$ galaxies is due to their later group/cluster infall times, such that they develop larger negative age differences prior to infall.

Massive ($M_\ast > 10^{10.5} \Msun$) spheroidal ($\kappaco < 0.35$) galaxies show almost no age differences across all environments.
This is potentially a result of the increasing contribution of mergers to the formation of massive galaxies \citep[e.g.][]{Rodriguez-Gomez_et_al_16, Clauwens_et_al_18, Davison_et_al_20}, which may act to erase any existing age gradients.

\subsection{Extent of recent star formation}

\begin{figure*}
  \includegraphics[width=\textwidth]{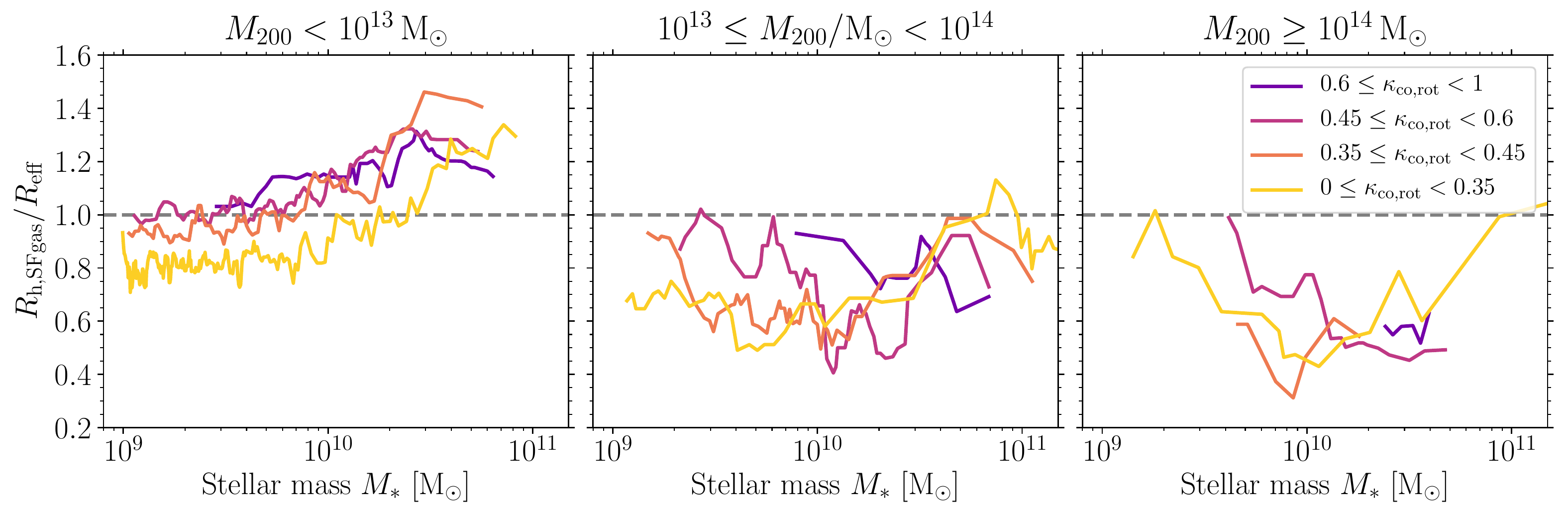}
  \caption{Half-mass radius of star-forming gas relative to $z=0$ galaxy effective radius, i.e. a measure of the concentration of the most recent star formation in the galaxies. Galaxies are divided into subsamples by $z=0$ environment and morphology as in Fig.~\ref{fig:dAge}. The trends of recent star formation extent follow those found for the inner/outer age differences in Fig.~\ref{fig:dAge}, indicating recent star formation in galaxies plays a major role in setting the age differences.}
  \label{fig:RhSFgas}
\end{figure*}

To investigate the cause of positive age gradients in group and cluster galaxies at late times, we first compare the distribution of star-forming gas in the galaxies.
In Fig.~\ref{fig:RhSFgas} we compare the median ratio of the half-mass radius of star-forming gas to the galaxy effective radius, for galaxies containing any star-forming gas. 
Star-forming gas is defined as that with a temperature within $0.5$~dex of temperature floor/equation of state imposed in EAGLE \citep[to prevent spurious fragmentation, see][for details]{Schaye_et_al_15}.

The trends found between the distribution of star-forming gas and galaxy mass, environment and morphology echo those found for inner/outer galaxy age differences in Fig.~\ref{fig:dAge}.
The star-forming gas distribution in massive, discy, field galaxies ($\kappaco > 0.35$, $M_\ast > 10^{10} \Msun$) is more extended than the stellar distributions as a whole, as expected from their negative age gradients (Fig~\ref{fig:dAge}).
Over time (and depending on the SFR) these galaxies would be expected to grow in size and develop negative age gradients.
In contrast, star-forming gas in group and cluster galaxies (of all types) is generally more compact than the galaxy size. 
Thus, group and cluster galaxies would be expected to develop positive age gradients (as found in Fig~\ref{fig:dAge}).
A similar correlation has been found for observed star-forming galaxies, where the ratio of the half-light radii for H$\alpha$ and continuum emission ($r_\mathrm{50,H\alpha} / r_\mathrm{50,cont}$) decreases in denser environments \citep{Schaefer_et_al_17}.

These results indicate that recent star formation plays a major role in the difference between field and cluster galaxies.
However, this difference may have a number of causes (e.g. late inner star formation peaks, outside-in quenching, etc.) which we discuss in the following sections.

\subsection{Inner star formation peaks}
\label{sec:SFpeaks}

\begin{figure*}
  \includegraphics[width=\textwidth]{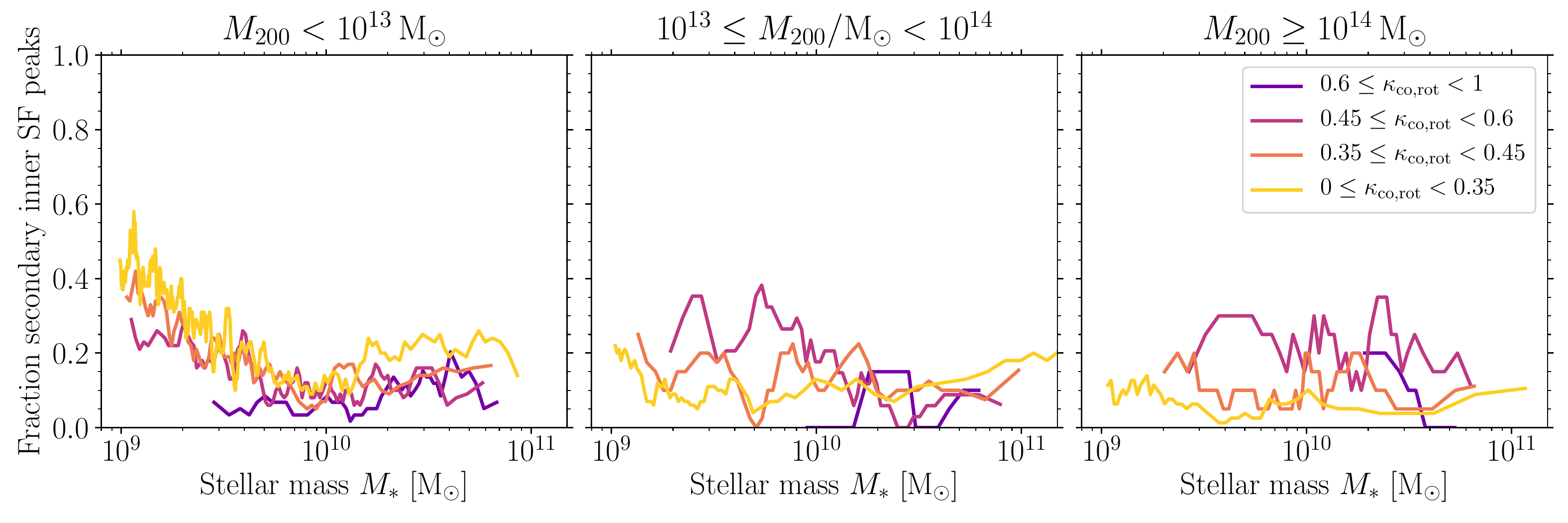}
  \caption{Fraction of galaxies that experience a secondary (or more) peak of star formation in the inner galaxy (within $\Reff$) with an amplitude at least 25 per cent of the maximum peak, and is not coincident with a peak in the outer galaxy (peaks offset by at least $1 \Gyr$). Generally less than 30 per cent of group and cluster galaxies have secondary star formation peaks, similar to field galaxies, suggesting younger inner regions of galaxies are generally not due to late, central star formation peaks.}
  \label{fig:SFpeaks}
\end{figure*}

To quantify whether the younger bulge ages (relative to the discs) of cluster galaxies is due to late, centrally concentrated episodes of star formation \citep[as proposed by][]{Johnston_et_al_14}, we determine the fraction of galaxies with secondary peaks in their star formation history.
We search for star formation peaks using the method described in Section~\ref{sec:peakFinding}.
Briefly, we define peaks in the star-formation histories where the maximum is $>50$ per cent above the previous minimum.

Fig.~\ref{fig:SFpeaks} shows fraction of galaxies with a significant secondary inner peak (peak amplitude at least 25 per cent of the maximum peak), where the peak happens at least $1 \Gyr$ after a significant peak in the outer galaxy (i.e. coincident inner/outer peaks are not counted).
Less than $\sim 30$ per cent of massive ($>10^{10} \Msun$ galaxies have secondary peaks, over all environments. Group and cluster galaxies also do not have significantly higher occurrences than field galaxies.
The results are not strongly sensitive to the exact fraction adopted, remaining less than $\sim$30 per cent for 5 and 10 per cent secondary peak amplitudes.

For low-mass galaxies ($< 10^{10} \Msun$) the fraction of secondary inner peaks depends somewhat on environment and morphology.
For spheroidal galaxies, secondary peaks are more common in field dwarf galaxies, with group and cluster galaxies having lower fractions. This may be due to star formation quenching following group/cluster infall, with low-mass cluster galaxies having typical infall redshifts of $z \sim 1$ (see Fig.~\ref{fig:zInfall}).
The strongest dependence on morphology also occurs in clusters, where more disc dominated (higher $\kappaco$) galaxies have a higher fraction of secondary peaks. Again, this may be caused by the later infall times of more disc-dominated galaxies (Fig.~\ref{fig:zInfall}), allowing them to remain star forming for longer.

Therefore in the EAGLE simulations, younger inner regions for group and cluster galaxies are, on average, not due to late star-formation peaks associated with quenching. Such an effect may therefore only be important in a subset of galaxies.
Instead, the inner regions of galaxies must have continued star formation over a longer period, relative to the outer regions.
In the following section, we discuss whether this effect may be purely environmental (e.g. through ram pressure stripping) or if the difference between field and group/cluster galaxies was in place prior to infall (progenitor bias).

\subsection{Redshift evolution}
\label{sec:zEvo}

\subsubsection{Age differences}

\begin{figure*}
  \includegraphics[width=\textwidth]{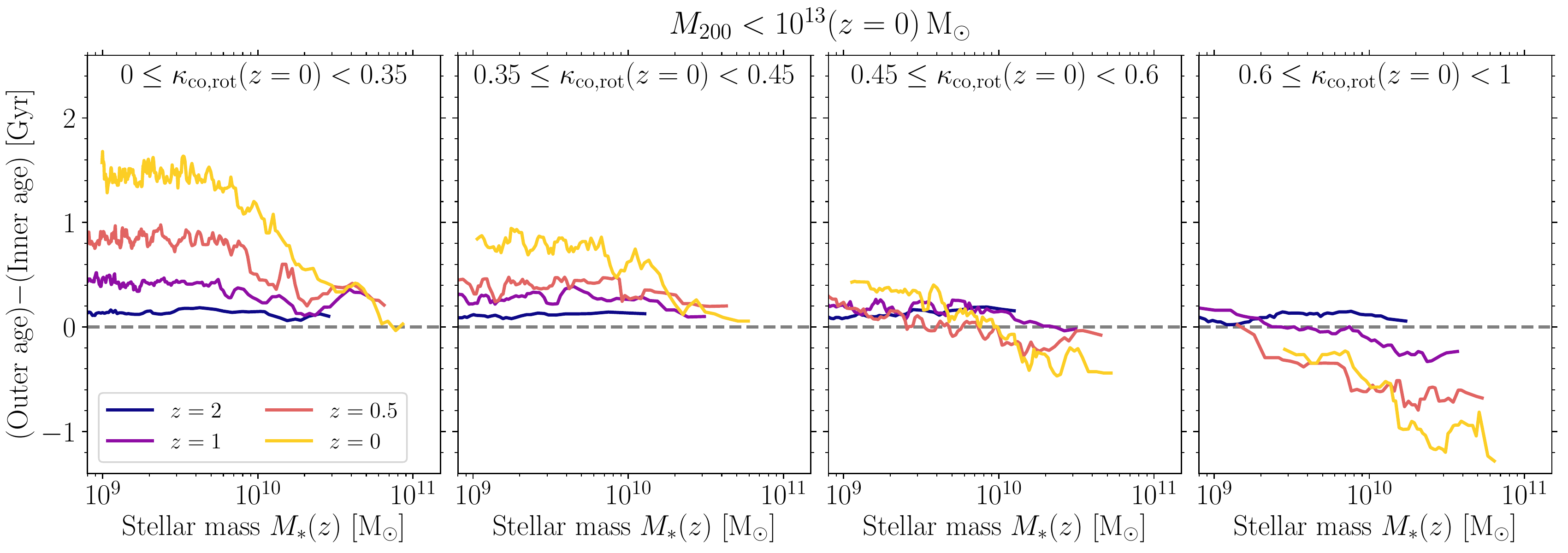}
  \includegraphics[width=\textwidth]{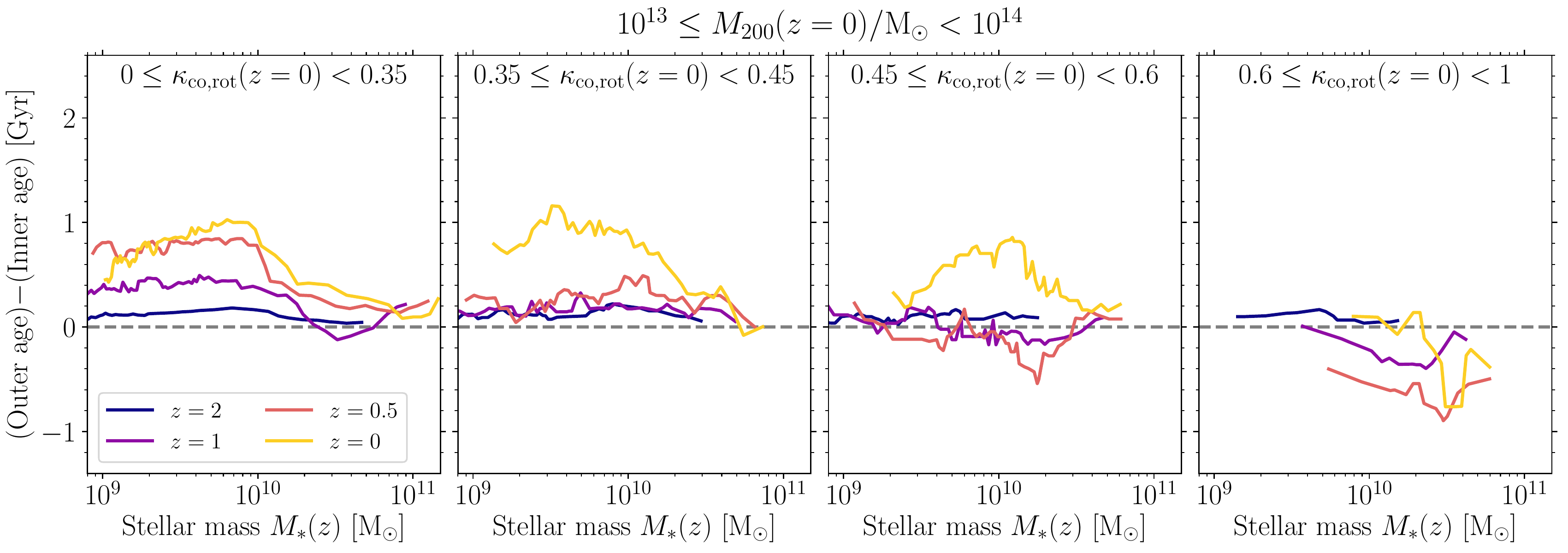}
  \includegraphics[width=\textwidth]{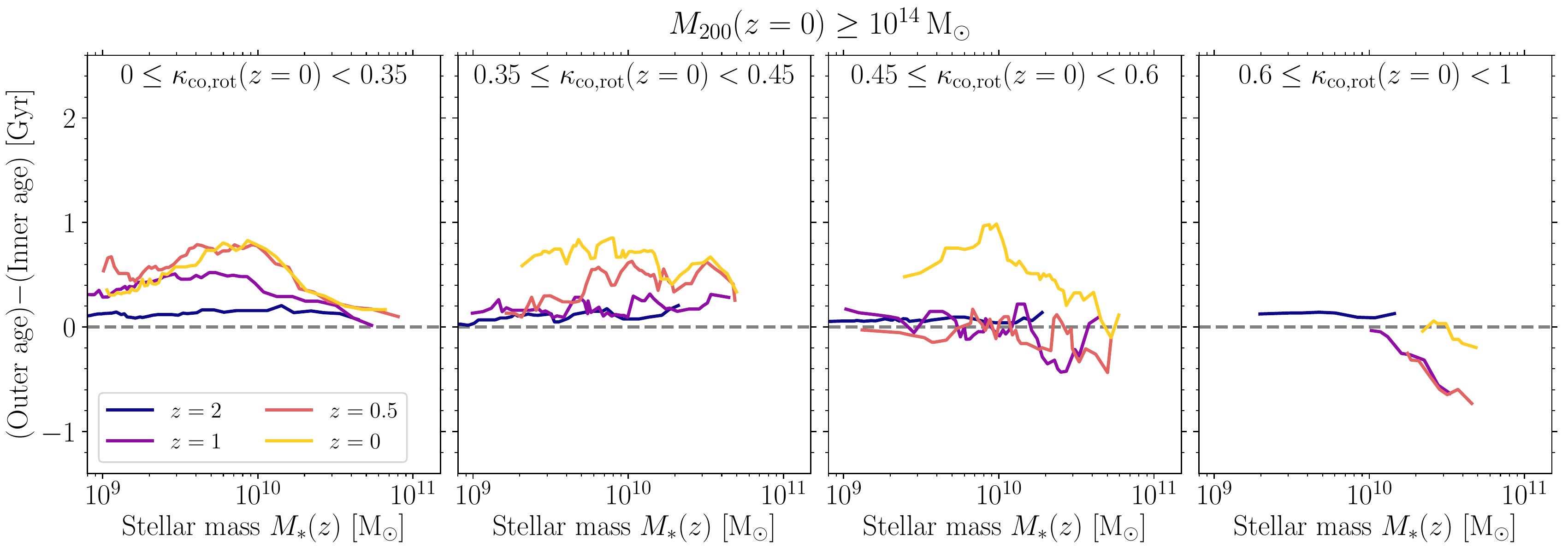}
  \caption{Redshift evolution of outer$-$inner age differences (relative to $\Reff$ at each redshift) for the main progenitors of galaxies in Fig.~\ref{fig:dAge}. The upper, middle and bottom rows of panels show `field', `group' and `cluster' galaxies (at $z=0$). In each row, $\kappaco$ of the galaxies at $z=0$ increases from left (spheroids) to right (very disc dominated). Line colours scale with the redshift of the progenitor galaxies. Progenitors of disc-dominated galaxies ($\kappaco > 0.45$) in groups and clusters (right columns, middle and bottom rows) begin to develop negative age differences by $z \sim 0.5$, with similar magnitudes to field galaxies (right columns, upper row), before being transformed into positive or near-zero age differences. The transformation time is related to the group/cluster infall time (see Fig.~\ref{fig:zInfall}).}
  \label{fig:dAge_evo}
\end{figure*}

If the progenitors of spiral/S0 galaxies in clusters were similar to present-day spirals, which then underwent further central star formation prior to quenching, we would expect the progenitors would initially show a negative age gradient (younger disc) before a reversal of the gradient after cluster infall.
We investigate this scenario in Figs.~\ref{fig:dAge_evo}, where we compare the age differences from redshifts $2$ to $0$ for the progenitors of galaxies from Fig.~\ref{fig:dAge}.
As previously, galaxies are divided by their $z=0$ morphology ($\kappaco$) and environment ($M_{200}$).

For field galaxies (upper row in Fig.~\ref{fig:dAge_evo}) the age gradients develop gradually. 
On average, the galaxies that will be more spheroid at $z=0$ ($\kappaco(z=0) < 0.45$) have age gradients that become more positive with time, while the inverse is true for the progenitors of disc-dominated galaxies ($\kappaco(z=0) > 0.45$).

For group and cluster galaxies the evolution of the age gradients is somewhat different.
For spheroidal galaxies ($\kappaco < 0.35$), the age gradients are largely set by $z \approx 0.5$, which is consistent with such galaxies being largely quenched by this time. 
For very discy galaxies ($\kappaco > 0.6$), they do indeed develop negative age gradients at $z=0.5$, before the gradients being largely erased by $z=0$.
For intermediate galaxies ($0.35 < \kappaco < 0.6$) in groups and clusters, they tend not to develop strong age gradients up to $z \approx 0.5$, before developing positive age gradients (younger inner regions) by $z=0$.

\begin{figure*}
  \includegraphics[width=\textwidth]{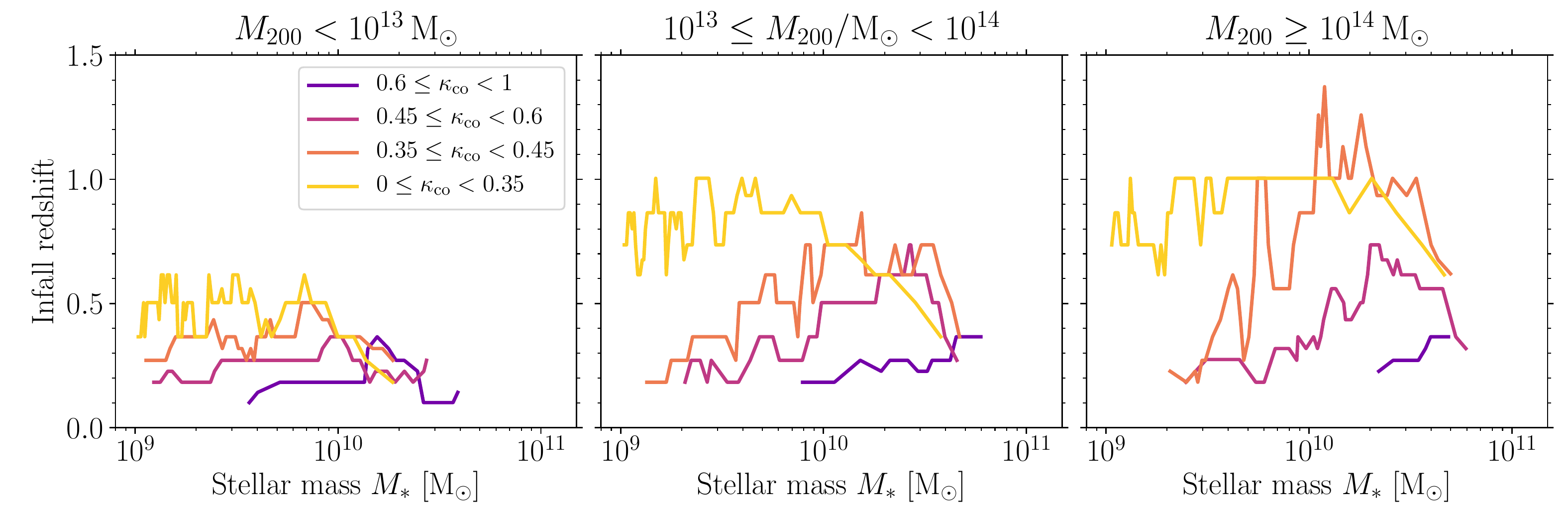}
  \caption{Infall redshifts of satellite galaxies. Panels and line colours are as in Fig.~\ref{fig:dAge}. Approximately 25, 93 and 99 per cent of the total galaxies are satellite galaxies in field, group and cluster environments, respectively (i.e. infall redshift if less relevant for field galaxies, where the majority are central galaxies of their halo). Infall redshift is correlated with both $\kappaco$ and $M_{200}$, such that galaxies with lower $\kappaco$ (more spheroidal) and higher $M_{200}$ (denser environments) have earlier infall times.}
  \label{fig:zInfall}
\end{figure*}

In Fig.~\ref{fig:zInfall} we show the median infall times of $z=0$ satellite galaxies (the latest snapshot at which the galaxy becomes a satellite subhalo in a FOF group), according to galaxy mass, environment and morphology as in Fig.~\ref{fig:dAge}.
There is a strong correlation between infall redshift and $\kappaco$, such that high $\kappaco$ galaxies are late infallers.
For $\kappaco > 0.45$ (disc-dominated galaxies) infall redshifts for groups/clusters occur at $z \lesssim 0.5$.
Indeed, at similar redshifts, galaxy clusters are observed to have a far higher proportion of blue and late-type galaxies than at the present day \citep{Butcher_and_Oemler_78, Butcher_and_Oemler_84, Couch_et_al_94, Couch_et_al_98}.
The time of the reversal of the age gradient for disc-dominated galaxies in groups/clusters (Fig.~\ref{fig:dAge_evo}) thus appears to be related to the infall times of the galaxies into the group/cluster environment.
At $z \approx 0.5$, the age differences (Fig.~\ref{fig:dAge_evo}) of group/cluster progenitors of disc galaxies were similar to that of the progenitors of field galaxies at the same redshift.
Therefore, some of the age gradient difference between field and cluster galaxies is a result a progenitor bias.
At the time of infall, cluster progenitor galaxies do not begin with the steeper age gradients of present-day field galaxies, but the flatter gradients of higher redshift galaxies.
For galaxies with very negative age gradients, such as disc-dominated galaxies at the present day, quenching may occur too quickly to fully reverse the age gradients.
The difference in evolution following group/cluster infall therefore likely has an environmental origin, which we investigate in the next section.

\subsubsection{Specific star formation profile evolution}
\label{sec:sSFR_evo}

\begin{figure*}
  \includegraphics[width=\textwidth]{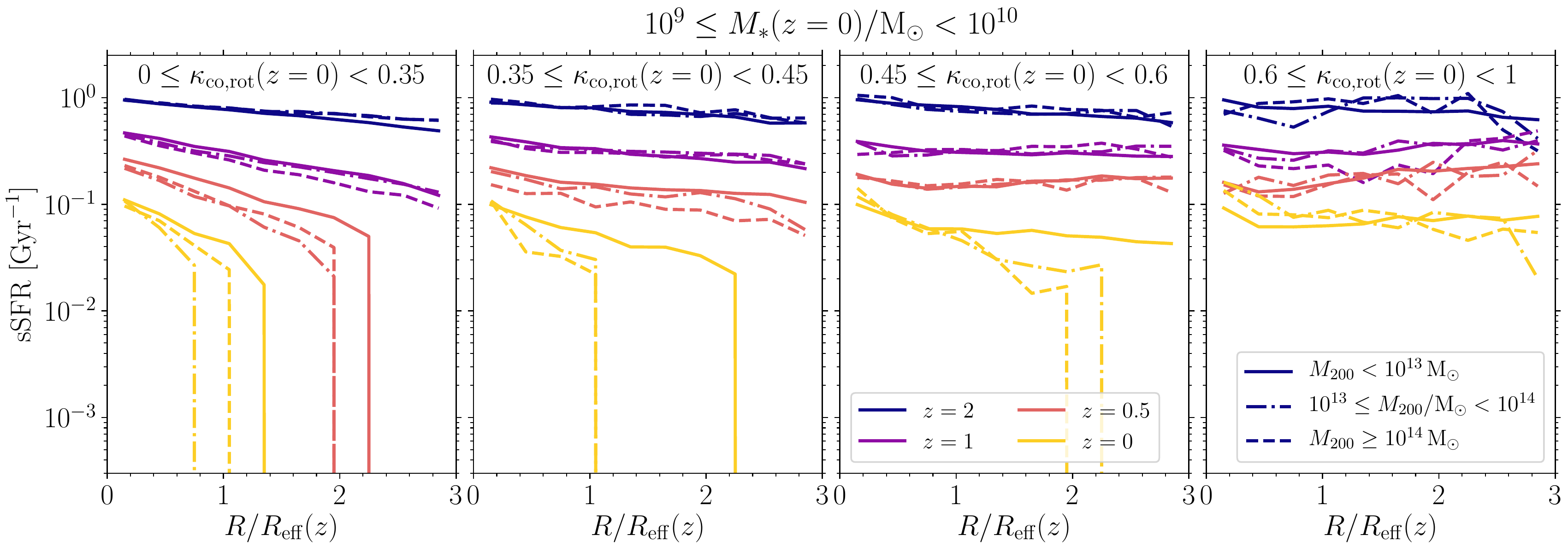}
  \includegraphics[width=\textwidth]{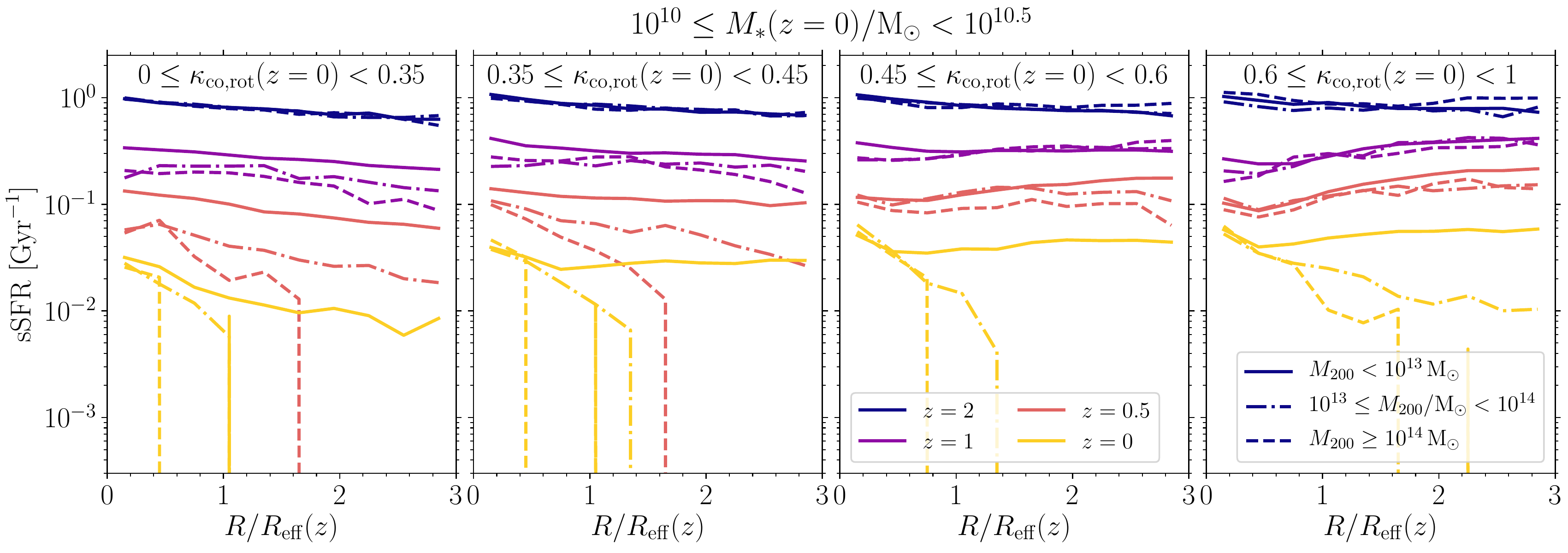}
  \includegraphics[width=\textwidth]{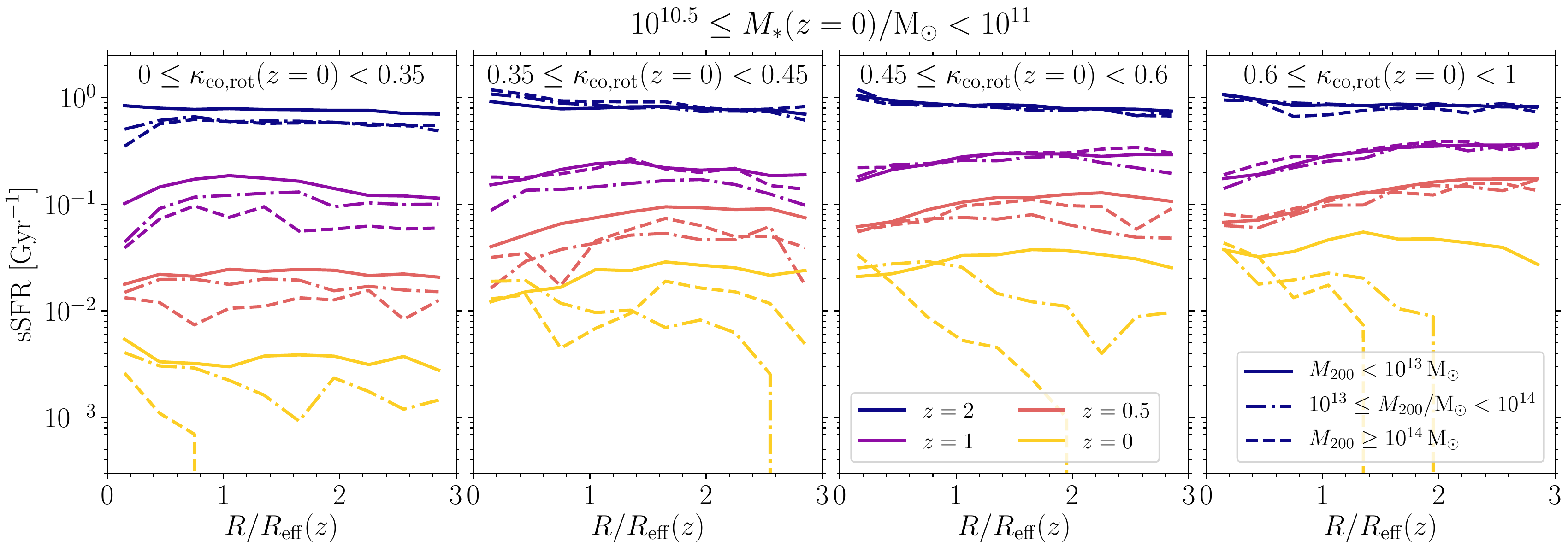}
  \caption{Redshift evolution of the median sSFR radial profiles for the main progenitors of galaxies in Fig.~\ref{fig:dAge}, limited to galaxies with $\mathrm{SFR}>0$. The upper, middle and bottom rows show the evolution of `low' ($10^9$-$10^{10} \Msun$), `intermediate' ($10^{10}$-$10^{10.5} \Msun$) and `high' ($10^{10.5}$-$10^{11} \Msun$) mass galaxies (mass at $z=0$), respectively. In each row, $\kappaco$ of the galaxies (at $z=0$) increases from left (spheroids) to right (very disc dominated). Solid, dash-dotted and dashed lines show galaxies from field, group and cluster environments, respectively, while line colour scales with progenitor redshift.}
  \label{fig:sSFR_evo}
\end{figure*}

We investigate the effect of environment on the star formation of galaxies by comparing the evolution of the radial sSFRs in Fig.~\ref{fig:sSFR_evo}.
Galaxies are first grouped by $z=0$ stellar mass (each row in the figure), $\kappaco$ at $z=0$ (panel in each row) and compared by environment in each panel (solid, dash-dotted and dotted lines for `field', `group' and `cluster', respectively).
Each colour shows the evolution of the sSFR profiles for the progenitor galaxies from $z=2$ to $z=0$.
The SFRs in each radial bin were calculated as the mass in stars formed within $300 \Myr$ at each snapshot, with the projected radii of star particles in each galaxy scaled by $\Reff$ of the progenitor galaxy at that redshift.
For each galaxy mass and morphology subgroup, we calculate the median sSFR for galaxies with $\mathrm{SFR} > 0$.

The sSFR profiles generally follow the profiles expected from the age differences (Fig.~\ref{fig:dAge_evo}).
At $z=2$ the profiles are near flat or slightly decreasing with increasing radius, consistent with the near zero age differences.
Near-flat sSFR profiles are similar to those found for observed galaxies at $z \sim 2$ \citep[e.g.][]{Liu_et_al_16, Liu_et_al_17, Wang_et_al_17}.
Low-mass, spheroidal galaxies ($10^9 \leq M_\ast/\mathrm{M}_{\sun} < 10^{10}$, $\kappaco < 0.35$) then develop radially decreasing sSFR profiles, while massive, disc-dominated field galaxies ($M_\ast > 10^{10} \Msun$, $\kappaco > 0.6$, $M_{200} < 10^{13} \Msun$) develop radially increasing profiles.

The sSFR profiles of progenitors of a given mass/morphology are similar between field and group/cluster galaxies prior to the typical infall redshift (Fig.~\ref{fig:zInfall}), explaining the similar age gradients in Fig.~\ref{fig:dAge_evo}.
For example, disc-dominated galaxies ($\kappaco > 0.45$) in all environments have similar sSFR profiles up to infall redshifts $z \lesssim 0.5$.
For massive spheroidal galaxies ($M_\ast > 10^{10} \Msun$, $\kappaco < 0.35$), the sSFR profiles for galaxies in group and cluster environments already begin to deviate by $z \approx 1$.

Post infall, the sSFR profiles for galaxies of a given mass and morphology then begin to differentiate by environment. 
In general, denser environments show sSFR profiles that are steeper and truncated at smaller galactocentric radii.
This difference is most drastic at $\gtrsim 1 \Reff$, where the sSFRs of group and cluster galaxies are suppressed relative to field galaxies (for all morphologies).
The central sSFRs ($\lesssim 0.5 \Reff$) instead remain relatively similar in all environments.
The cold gas consumption times ($M_\mathrm{g,cold} / \mathrm{SFR}$) is typically 2-3 Gyr for these galaxies, even those with suppressed outer sSFRs, meaning the centres of the galaxies may remain star forming for a significant time without further gas removal.
Thus, the galaxies in group/cluster environments are being quenched from outside-in, which explains their younger centres (positive age gradients) relative to field galaxies of a similar morphology.

\section{Discussion}
\label{sec:discussion}

\subsection{Transformation of cluster galaxies}

The different age gradients of field and cluster disc galaxies \citep[both spirals and S0s, e.g.][]{Roediger_et_al_11, Johnston_et_al_14} suggests different formation paths or transformation processes are at play for different environments.
Analysing the EAGLE simulations, we find that ISM stripping is likely the key driver of these age differences. Outside-in quenching results in centrally concentrated star formation prior to complete quenching of the cluster galaxies, explaining the younger centres of the cluster galaxies.
The results in this work thus support the hypothesis that spiral galaxies are transformed into S0 galaxies in dense environments through outside-in quenching \citep[e.g.][]{Bedregal_et_al_11}.

The decreased outer sSFRs of cluster galaxies in the simulations also agree with the observations of cluster galaxies.
Relative to field galaxies, Virgo cluster spiral galaxies have truncated H$\alpha$ discs and outer SFRs decreased by factors of $1.5$-$9$, while inner SFRs are similar or enhanced up to a factor of $1.7$, suggesting ISM stripping as the most likely origin \citep{Koopmann_and_Kenney_04a, Koopmann_and_Kenney_04b}.
\citet{Schaefer_et_al_17} also found that H$\alpha$ emission becomes more concentrated (relative to the continuum emission) for galaxies in denser environments.
Recently, \citet{Matharu_et_al_21} found evidence for outside-in quenching of cluster galaxies at $z \sim 1$. They found that recently quenched cluster galaxies have significantly more concentrated H$\alpha$ emission compared to star-forming galaxies of similar mass, and similarly suggest ram-pressure stripping as the likely cause.

We find this transformation to centrally concentrated star formation for cluster galaxies is (on average) coincident with their infall times (first time the galaxies become a satellite in a group/cluster; Fig.~\ref{fig:zInfall}), typically occurring at redshifts $z \lesssim 0.5$ for disc-dominated galaxies (with the most disc-dominated galaxies at $z=0$ having the latest infall times).
Observations of galaxy clusters at similar redshifts show they have a far higher proportion of spiral galaxies than clusters at the present day \citep{Couch_et_al_94, Couch_et_al_98, Fasano_et_al_00, Desai_et_al_07}, consistent with recent infall prior to transformation into S0 galaxies at later redshifts.

In such distant galaxy clusters ($z \approx 0.3$) approximately 30 per cent of galaxies are found to be undergoing secondary episodes of star formation \citep{Couch_and_Sharples_87, Barger_et_al_96}, which appear to be systems involved in mergers \citep{Couch_et_al_98}.
In Section~\ref{sec:SFpeaks} we find $< 30$ per cent of cluster galaxies have undergone secondary \textit{central} episode of star formation.
This does not necessarily imply tension between observations and simulations given the very different methods used, and that we exclude any galaxies that undergo galaxy-wide or only outer secondary episodes (the fraction of galaxies with secondary episodes of star formation of course increases when relaxing these restrictions, by a factor $\approx 2$ on average).
Future work should use consistent methods to determine if the fraction of galaxies in high-redshift clusters with secondary star formation episodes can be explained in large-scale hydrodynamical simulations.

\subsection{Limitations of the EAGLE model}

Predictions for the age gradients of galaxies may of course depend upon the details of the simulation being analysed.
Particularly relevant for this work, the central SFRs of simulated galaxies have been shown to depend upon the implementation of AGN feedback \citep{Appleby_et_al_20, Nelson_et_al_21}.

\citet{Lagos_et_al_20} found low velocity dispersions in centres of massive ($M_\ast > 10^{10} \Msun$) EAGLE galaxies, which they find is related to young stellar populations.
This may also be connected to the too centrally-concentrated sSFR profiles found for EAGLE galaxies below the main star-forming sequence (``green valley'' galaxies) \citep{Starkenburg_et_al_19}.
These findings suggest, although effective on a galaxy-wide scale, AGN feedback in EAGLE may be insufficient to quench the central star formation in massive galaxies.
We also note that the density profiles of massive EAGLE galaxies are insufficiently concentrated \citep[i.e. deficient in stellar mass near the centres of the galaxies, potentially due to spatial resolution and cooling floor limits,][]{de_Graaff_et_al_21}, which may exacerbate the discrepancy in sSFR at galactic centres.

In Appendix~\ref{app:variations} (Fig.~\ref{fig:AGNdT9-dAge}) we investigated an EAGLE model variation which implements a higher AGN heating temperature (AGNdT9-L050N0752, the model provides a better match to X-ray observations of intragroup and intracluster gas) in order to test whether more effective AGN feedback affects the resulting age gradients.
However, we find very similar age differences to the EAGLE reference model (Fig.~\ref{fig:dAge}).
How these findings of insufficient central quenching may affect the age gradient predictions is not clear (e.g. whether this implies an offset to younger outer regions for all galaxy types, only green valley galaxies, or predominately the more spheroidal galaxies which tend to have more centrally concentrated star formation, Fig.~\ref{fig:sSFR_evo}).

\section{Summary}
\label{sec:summary}

In this work we used the EAGLE simulations to investigate the origin of the positive age gradients in cluster spiral and S0 galaxies.
We found that massive ($M_\ast > 10^{10} \Msun$), disc-dominated ($\kappaco > 0.45$), field galaxies ($M_{200} < 10^{13} \Msun$) in EAGLE have younger outer ($>\Reff$; `disc') regions than inner regions on average, while similar galaxies in group and cluster environments ($M_{200} > 10^{13} \Msun$) tend to have similarly aged or younger inner ($<\Reff$; `bulge') regions relative to their outer regions (Fig.~\ref{fig:dAge}).
This result is qualitatively similar to the age gradients found for observed spiral and S0 galaxies \citep[e.g.][]{Roediger_et_al_11, Gonzalez_Delgado_et_al_14, Johnston_et_al_14, Fraser-McKelvie_et_al_18}.

Lower mass galaxies in the simulations ($M_\ast < 10^{10} \Msun$) generally show near-zero or positive age differences (Fig.~\ref{fig:dAge}). Again, this is qualitatively consistent with that found for observed galaxies \citep{Perez_et_al_13, Fraser-McKelvie_et_al_18, Breda_et_al_20}.
\citet{Graus_et_al_19} find this is due to feedback in dwarf galaxies pushing old stars to larger radii \citep[though spurious collisional heating may also play some role,][]{Ludlow_et_al_19, Ludlow_et_al_21}.

We used the simulations to test the origins of the difference in age gradients of field and group/cluster galaxies: outside-in quenching in denser environments, late episodes of star formation in galaxy bulges and galaxy progenitor bias (where field progenitors of cluster galaxies are markedly different from present-day field galaxies).
In Section~\ref{sec:SFpeaks} we show that late, central star formation episodes in cluster galaxies play a more minor role, typically affecting $<30$ per cent of galaxy populations. The proportion of galaxies with such secondary episodes is also not significantly different from the general field population.

We found that the younger inner regions (compared to their outer regions) of cluster disc galaxies, relative to field galaxies, are a result of both progenitor bias and outside-in quenching (Section~\ref{sec:zEvo}).
Both field and group/cluster progenitor galaxies have similar age and sSFR gradients up to the time of group/cluster infall, after which their evolution diverges.
The outer sSFRs of group/cluster galaxies significantly decrease while the central sSFRs (for those galaxies which are not yet fully quenched of star formation) are similar to field galaxies.
From the flatter age gradients of higher redshift galaxies (i.e. galaxy progenitor bias), the group/cluster galaxies then develop positive age gradients due to the ongoing central star formation.
Field disc galaxies instead continue to develop stronger negative age gradients.
Outside-in quenching of galaxies in groups/clusters is consistent with the expected effects from stripping of the interstellar medium \citep[e.g.][]{Koopmann_and_Kenney_04b, Bekki_09, Owers_et_al_19}.

\section*{Acknowledgements}

We thank Amelia Fraser-McKelvie, Luca Cortese and Claudia Lagos for helpful comments and discussion.
This research was supported by the Australian government through the Australian Research Council’s Discovery Projects funding scheme (DP200102574).
This work used the DiRAC Data Centric system at Durham University, operated by the Institute for Computational Cosmology on behalf of the STFC DiRAC HPC Facility (\url{www.dirac.ac.uk}). This equipment was funded by BIS National E-infrastructure capital grant ST/K00042X/1, STFC capital grants ST/H008519/1 and ST/K00087X/1, STFC DiRAC Operations grant ST/K003267/1 and Durham University. DiRAC is part of the National E-Infrastructure.

\section*{Data Availability}

All data (including galaxy catalogues, merger trees and particle data) from the EAGLE simulations is publicly available \citep{McAlpine_et_al_16} at \href{http://www.eaglesim.org/database.php}{http://www.eaglesim.org/database.php}.


\bibliographystyle{mnras}
\bibliography{bibliography}



\appendix

\section{Metallicity and colour differences}
\label{app:colours}

\begin{figure*}
  \includegraphics[width=\textwidth]{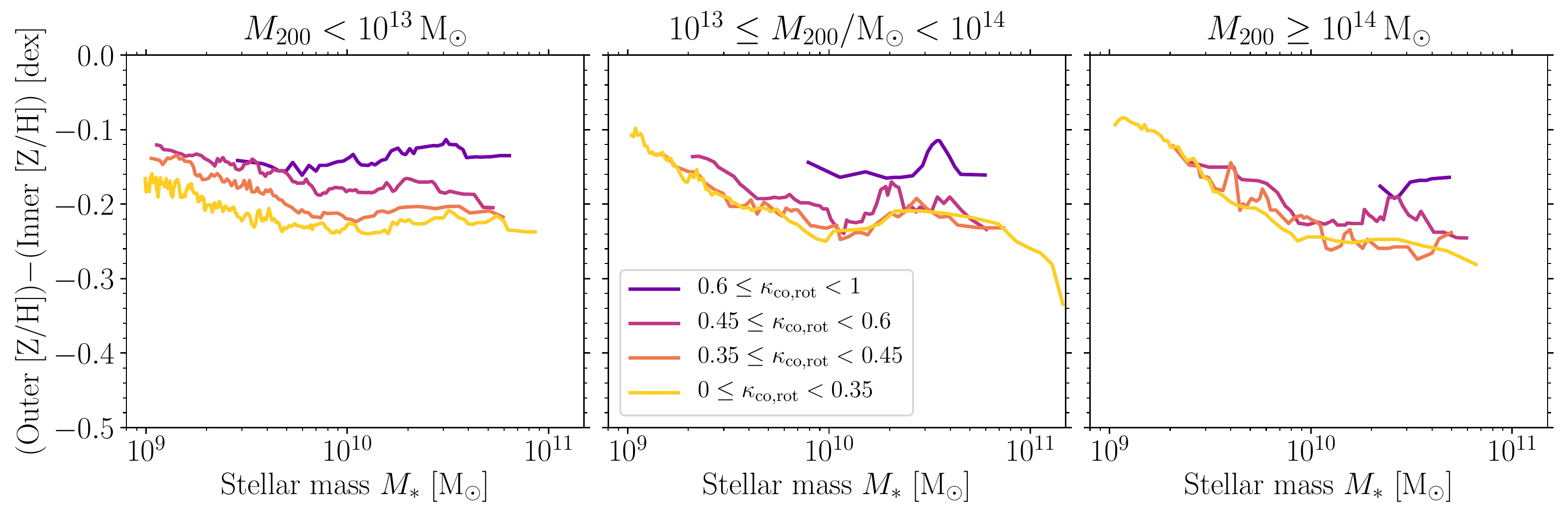}
  \caption{Outer$-$inner metallicity differences for EAGLE galaxies as a function of galaxy mass, environment (panels, increasing $M_{200}$ left to right) and morphology ($\kappaco$, lines coloured according to legend). The lines show running medians as a function of galaxy mass, as in Fig~\ref{fig:dAge}.}
  \label{fig:dZH}
\end{figure*}

For reference, in this section we compare the outer$-$inner metallicity and colour differences for EAGLE galaxies as a function of galaxy mass, environment and morphology (as for the age differences in Fig.~\ref{fig:dAge}).

In Fig.~\ref{fig:dZH} we compare the outer$-$inner metallicity differences.
All galaxy subsamples show negative metallicity gradients on average, with higher $\kappaco$ (more disc dominated) galaxies showing the smallest metallicity differences.
For field galaxies, the metallicity difference depends mainly on $\kappaco$ and little on galaxy mass.
Group and cluster galaxies instead have a stronger dependence on galaxy mass (with higher mass galaxies showing larger metallicity differences), with less dependence on $\kappaco$ than field galaxies.
This metallicity difference naturally follows the age differences found in Fig.~\ref{fig:dAge}, such that, relative to the inner regions of galaxies, the youngest outer regions have experienced the most self-enrichment in metallicity.
The typical metallicity differences of $<0.3$~dex are similar to those found in observed galaxies using a range of stellar population modelling techniques (e.g. fitting broadband colours, \citealt{Roediger_et_al_11}; Lick indices, \citealt{Fraser-McKelvie_et_al_18}; and full spectral fitting, \citealt{Breda_and_Papaderos_18}.

\begin{figure*}
  \includegraphics[width=\textwidth]{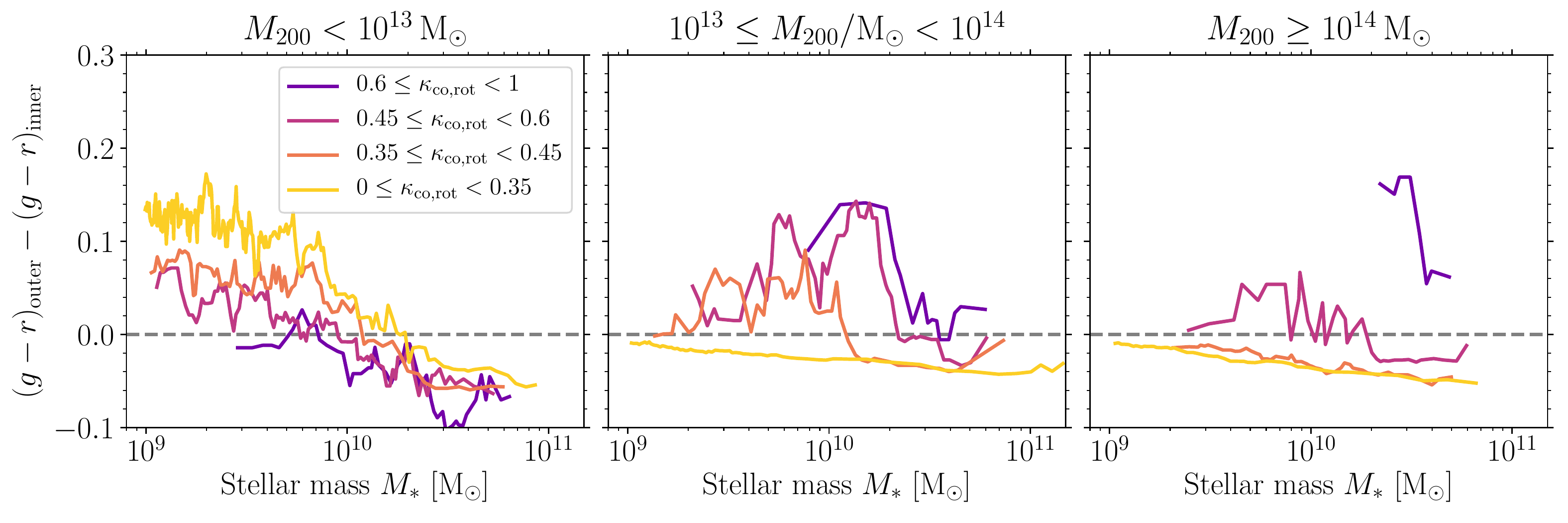}
  \includegraphics[width=\textwidth]{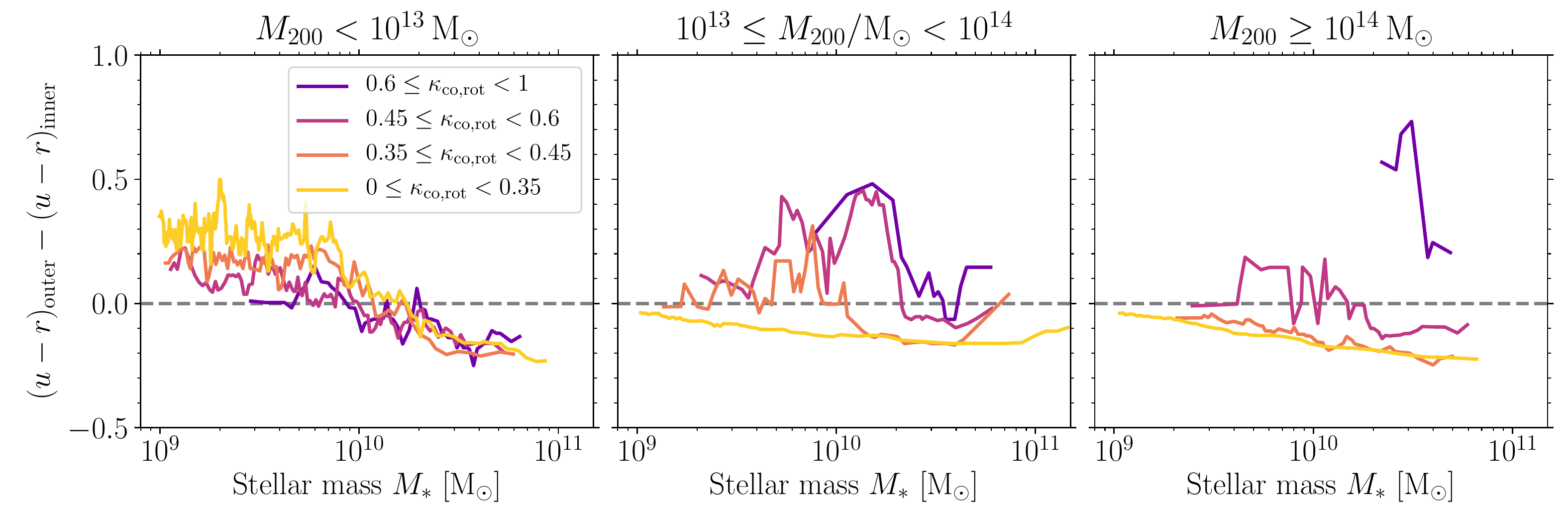}
  \includegraphics[width=\textwidth]{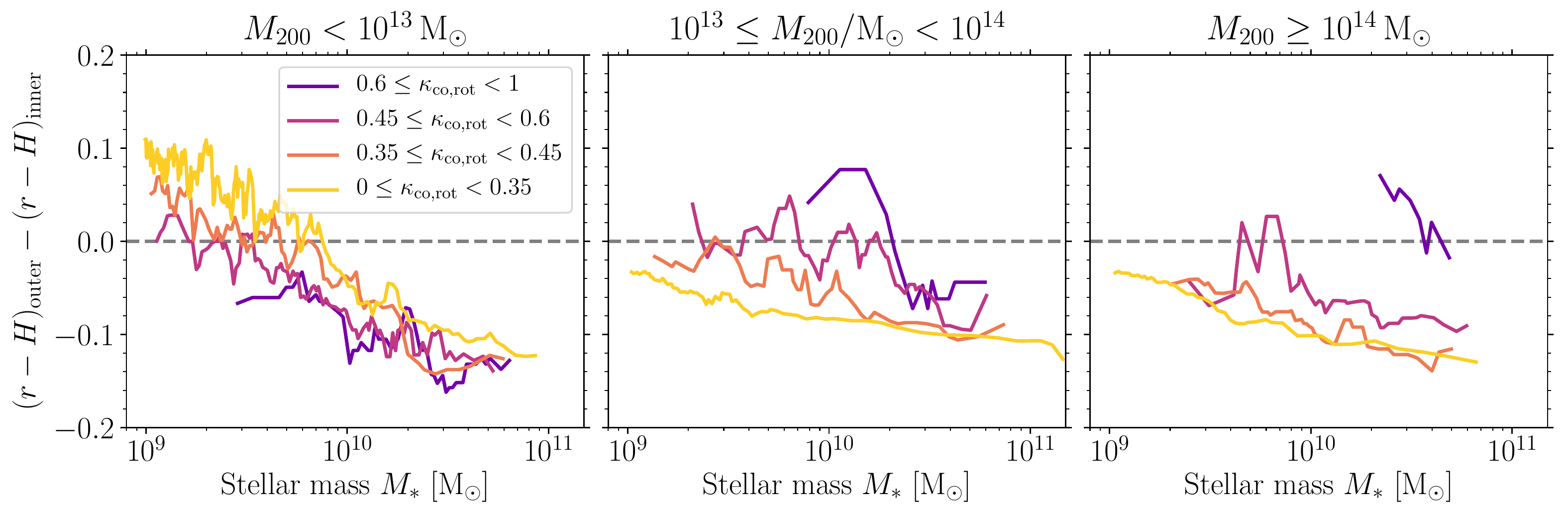}
  \caption{Outer$-$inner colour differences for EAGLE galaxies as a function of galaxy mass, environment (panels, increasing $M_{200}$ left to right) and morphology ($\kappaco$, lines coloured according to legend). The lines show running medians as a function of galaxy mass, as in Fig.~\ref{fig:dAge}.}
  \label{fig:colours}
\end{figure*}

Fig.~\ref{fig:colours} compares some commonly investigated colours in the literature \citep[e.g.][]{Prochaska_Chamberlain_et_al_11, Mishra_et_al_17}, namely $g-r$, $u-r$ and $r-H$ (for SDSS bands $u$, $g$ and $r$, 2MASS $H$).
For field galaxies ($M_{200} < 10^{13} \Msun$), low mass galaxies ($M_\ast <10^{10} \Msun$) tend to have redder outer regions, consistent with their positive age differences in Fig.~\ref{fig:dAge}, with less dependence on morphology ($\kappaco$).
Group and cluster galaxies ($M_{200} > 10^{13} \Msun$) have a stronger dependence on morphology, with more spheroidal galaxies showing more negative colour differences (bluer outer regions).
This stems from the smaller absolute age differences between high and low mass galaxies in group and cluster environments compared to field galaxies.

\section{EAGLE model variations}
\label{app:variations}

\begin{figure*}
  \includegraphics[width=\textwidth]{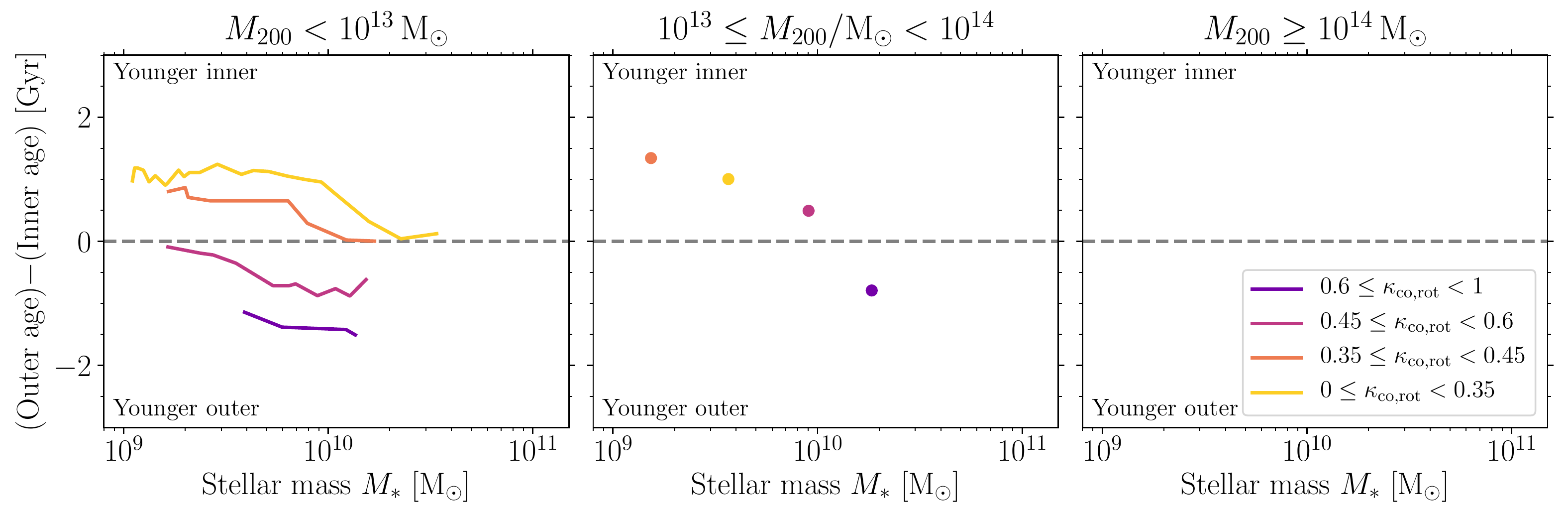}
  \caption{As for Fig.~\ref{fig:dAge}, but comparing the age differences in the EAGLE high resolution model Recal-L025N0752. Given the smaller simulation volume ($25^3$ comoving Mpc$^3$), there are no galaxy clusters ($M_{200} \geq 10^{14} \Msun$) and groups ($10^{13} \leq M_{200} / \mathrm{M}_{\sun} < 10{14}$) are poorly sampled. We show only single points where there are too few galaxies to obtain a running median ($<25$). The results are generally consistent with the EAGLE reference model, showing only slightly larger (more negative) age differences for disc-dominated galaxies ($\kappaco > 0.45$).}
  \label{fig:RECAL-dAge}
\end{figure*}

\begin{figure*}
  \includegraphics[width=\textwidth]{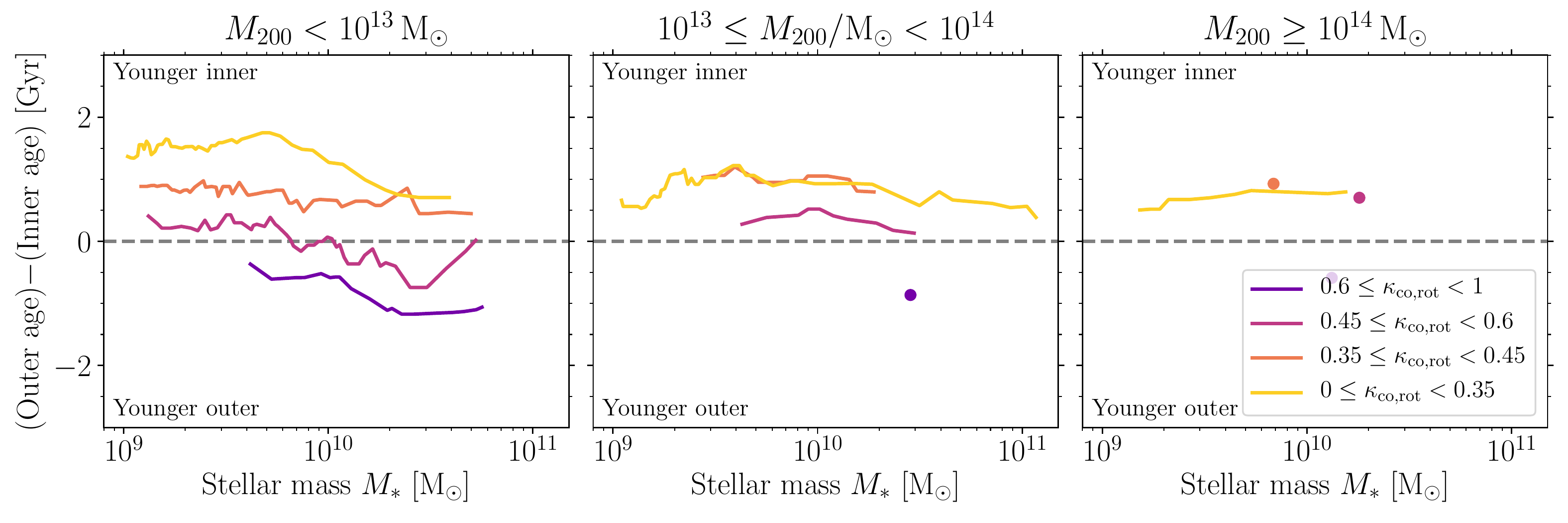}
  \caption{As for Fig.~\ref{fig:dAge}, but comparing the age differences in the EAGLE model variation AGNdT9-L050N0752, which implements a higher temperature for AGN feedback ($10^9 \K$) relative the the reference model ($10^{8.5} \K$). We do not find significant differences from the EAGLE reference model.}
  \label{fig:AGNdT9-dAge}
\end{figure*}

In this section we compare the age difference predictions (Section~\ref{sec:dAge} and Fig.~\ref{fig:dAge}) for two different EAGLE model variations, Recal-L025N0752 and AGNdT9-L050N0752 \citep{Schaye_et_al_15}.

Recal-L025N0752 has 8 times higher mass resolution ($m_g = 2.25 \times 10^5 \Msun$) and 2 times small gravitational softening (maximum $0.35$ proper kpc) compared to the reference model, and uses recalibrated stellar and AGN feedback parameters to better match the $z \sim 0$ galaxy stellar mass function at this resolution. The periodic volume has a side length of 25 comoving Mpc, and thus a volume 64 times smaller than the reference mode.

AGNdT9-L050N0752 uses the same resolution as reference model, but with slightly modified AGN parameters (in particular a higher heating temperature of $\Delta T_\mathrm{AGN} = 10^9 \K$, compared to $10^{8.5} \K$) to better match X-ray observations of intragroup and intracluster gas.
The periodic volume has a side length of 50 comoving Mpc, and thus a volume 8 times smaller than the reference mode.

Figures \ref{fig:RECAL-dAge} and \ref{fig:AGNdT9-dAge} compare the age differences for both models, computed in an identical manner as the reference model in Fig.~\ref{fig:dAge}.
We find the results are consistent between all three models.
The Ref-L100N1504 and AGNdT9-L050N0752 models show almost identical age differences as a function of mass, morphology and environment.
Compared to the other models, Recal-L025N0752 shows slightly younger outer regions for disc-dominated galaxies ($\kappaco > 0.45$), which may be related to the slightly higher low-redshift ($z \sim 0$) specific star formation rates \citep{Schaye_et_al_15, Furlong_et_al_15}, assuming these higher SFRs are associated with outer disc formation.


\bsp	
\label{lastpage}
\end{document}